\documentclass[11pt,superscriptaddress,aps,prd,preprint]{revtex4}
\usepackage[dvips]{graphicx}
\usepackage{amsfonts}

\newcommand{\Slash}[1]{\!\not\!{#1}}

\def\be{\begin{equation}}
\def\en{\end{equation}}

\def\ba{\begin{eqnarray}}
\def\ea{\end{eqnarray}}

\def\tr{\mathrm{tr}}

\def\sumint{\hbox{$\sum$}\!\!\!\!\!\!\int}

\begin{document}

\title{Free energy of Lorentz-violating QED at high temperature}

\author{M. Gomes}
\affiliation{Instituto de F\'{\i}sica, Universidade de S\~ao Paulo\\
Caixa Postal 66318, 05315-970, S\~ao Paulo, SP, Brazil}
\email{mgomes,ajsilva@fma.if.usp.br}

\author{T. Mariz}
\affiliation{Instituto de F\'\i sica, Universidade Federal de Alagoas, 57072-270, Macei\'o, Alagoas, Brazil}
\email{tmariz@if.ufal.br}

\author{J. R. Nascimento}
\affiliation{Departamento de F\'{\i}sica, Universidade Federal da Para\'{\i}ba\\
Caixa Postal 5008, 58051-970, Jo\~ao Pessoa, Para\'{\i}ba, Brazil}
\email{alesandro,jroberto,petrov@fisica.ufpb.br}

\author{A. Yu. Petrov}
\affiliation{Departamento de F\'{\i}sica, Universidade Federal da Para\'{\i}ba\\
Caixa Postal 5008, 58051-970, Jo\~ao Pessoa, Para\'{\i}ba, Brazil}
\email{alesandro,jroberto,petrov@fisica.ufpb.br}

\author{A. F. Santos}
\affiliation{Departamento de F\'{\i}sica, Universidade Federal da Para\'{\i}ba\\
Caixa Postal 5008, 58051-970, Jo\~ao Pessoa, Para\'{\i}ba, Brazil}
\email{alesandro,jroberto,petrov@fisica.ufpb.br}

\author{A. J. da Silva}
\affiliation{Instituto de F\'{\i}sica, Universidade de S\~ao Paulo\\
Caixa Postal 66318, 05315-970, S\~ao Paulo, SP, Brazil}
\email{mgomes,ajsilva@fma.if.usp.br}

\begin{abstract}
In this paper we study the one- and two-loop contribution to the free energy in QED with the Lorentz symmetry breaking introduced via constant CPT-even Lorentz-breaking parameters at the high temperature limit. We find the impact of the Lorentz-violating term for the free energy and carry out a numerical estimation for the Lorentz-breaking parameter.
\end{abstract}

\maketitle

\section{Introduction}

Nowadays the Lorentz symmetry breaking is treated as an important ingredient of field theory models for probing quantum gravity phenomena. In fact, Lorentz-violating theories have been studied in various contexts, such as string theory \cite{Kos}, noncommutative field theory \cite{Car}, and, more recently, Horava-Lifshitz gravity \cite{Hor}.

Many of the effects observed for the Lorentz-breaking field theories at zero temperature are known to persist also at finite temperatures. For example, the dependence of the loop corrections on the regularization scheme in the finite temperature case was shown to arise in the one-loop order in the Lorentz-breaking quantum electrodynamics (QED) \cite{bnpp}. Analogous situation takes place in the Lorentz-breaking Yang-Mills theory \cite{YM4d}. Different issues related to the Lorentz-violating QED in the finite temperature case were considered in \cite{others}. In \cite{Cas} the finite temperature properties in the CPT-odd Lorentz-breaking extension of QED for a purely spacelike background were studied, and in \cite{Cas1} these properties were analysed for the CPT-even Lorentz-breaking extension.

Most of the previous studies of the Lorentz-breaking theories, including those described in \cite{bnpp,YM4d}, were based on the use of couplings which violate not only Lorentz symmetry but also the CPT symmetry, although CPT-even Lorentz breaking interactions are also possible and certainly require more detailed study. In this paper, we will see that, at the high temperature regime, the linear contribution in the Lorentz-breaking parameter may arise.

The aim of this paper is the study of one and two-loop corrections to the free energy (second order in the coupling constant $e$) of QED in the presence of Lorentz breaking terms, at finite temperature. As it is known the free energy provides an important information to different physical issues such as plasma behavior, solar interior and Big Bang nucleosynthesis (BBN) (see, \cite{Dicus,Heck,Lopez}). In this context, some aspects of the corrections to the free energy in theories without Lorentz-breaking are discussed in \cite{And,Kapusta}. 

For estimating bounds of the Lorentz-breaking parameter, we use information of the BBN (for a review, see \cite{Ioc}), which is one of the observational pillars of the standard cosmology. Note that the Lorentz-violating parameter can explain for the light elements abundance. In particular, the difference between the theoretical and observational results can be understood as come from a contribution of the Lorentz-breaking parameter to the primordial helium abundance.

The structure of the paper is as follows. In Sec.~\ref{lvQED} we present the basic features of the Lorentz-violating QED in the regime of high temperature. In Sec.~\ref{fe} we calculate the one and two-loop contributions to the free energy for QED involving the Lorentz-breaking fermion coupling. In Sec.~\ref{ne}, by using information about the primordial helium abundance, we obtain a numerical estimation for the Lorentz violation parameter. A summary is presented in Sec.~\ref{su}.

\section{Lorentz-violating QED at high temperature}\label{lvQED}

Let us start by considering the Lagrangian of the Lorentz- and CPT-violating QED extension \cite{SME}
\be\label{Lbfe}
{\cal L}=-\frac{1}{4}F_{\mu\nu}F^{\mu\nu}-\frac14 (k_F)_{\mu\nu\lambda\rho}F^{\mu\nu}F^{\lambda\rho}+\frac12 (k_{AF})^\mu\epsilon_{\mu\nu\lambda\rho}A^\nu F^{\lambda\rho}+\bar{\psi}\left(i\Gamma^\mu D_\mu-M \right)\psi + {\cal L}_\mathrm{gf} + {\cal L}_\mathrm{gh},
\en
where $\Gamma^\mu = \gamma^\mu+\Gamma_1^\mu$, $M=m+M_1$, with
\ba
\Gamma_1^\mu &=& c^{\mu\nu}\gamma_\nu + d^{\mu\nu}\gamma_5\gamma_\nu + e^\mu + if^\mu\gamma_5 + \textstyle{1\over 2}g^{\lambda\nu\mu}\sigma_{\lambda\nu} \\
M_1 &=& a_\mu\gamma^\mu + b_\mu\gamma_5\gamma^\mu + \textstyle{1\over 2}H_{\mu\nu}\sigma^{\mu\nu},
\ea
and $D_\mu = \partial_\mu + ieA_\mu$. ${\cal L}_\mathrm{gf}$ is the gauge fixing term and ${\cal L}_\mathrm{gh}$ is the ghost field term, which decouples from the rest of the Lagrangian. The coefficients carrying an odd (even) number of Lorentz indices are CPT-odd (-even).  

The two Lorentz-violating terms of the photon sector, the Chern-Simons-like (CPT-odd) and $(k_F)_{\mu\nu\lambda\rho}F^{\mu\nu}F^{\lambda\rho}$ (CPT-even) terms, can be induced by radiative corrections from the terms with the coefficients $b_\mu$ and $c_{\mu\nu}$ of the fermion sector, respectively, so that $(k_{AF})_\mu\propto b_\mu$ \cite{JacKos} and \cite{KosLan} 
\be\label{kF}
(k_F)_{\mu\nu\lambda\rho} \propto \frac12 g_{\mu\lambda}(c_{\nu\rho}+c_{\rho\nu})+\frac12 g_{\nu\rho}(c_{\mu\lambda}+c_{\lambda\mu})-\frac12 g_{\mu\rho}(c_{\nu\lambda}+c_{\lambda\nu})-\frac12 g_{\nu\lambda}(c_{\mu\rho}+c_{\rho\mu}).
\en

The coefficients $a_\mu$, $b_\mu$, $H_{\mu\nu}$, and $(k_{AF})_\mu$ have dimensions of mass, while $c_{\mu\nu}$, $d_{\mu\nu}$, $e_\mu$, $f_\mu$, $g_{\lambda\nu\mu}$, and $(k_F)_{\mu\nu\lambda\rho}$ are dimensionless. In the high temperature regime ($T\gg M$) the dimensionful coefficients may be neglected. For instance, the $b_\mu$-corrections to the free energy are observed to be proportional to $b^2T^2$, similarly to the scenario occurring with the corrections stemming from the Chern-Simions-like term \cite{Cas}. Thus both $b_\mu$ and $(k_{AF})_\mu$ are negligible at high temperature, as well as  $a_\mu$ and $H_{\mu\nu}$.

Among the dimensionless coefficients, $e_\mu$, $f_\mu$, and $g_{\lambda\nu\mu}$ are expected to be much smaller than the other, because their terms cannot be obtained directly from the standard model extension \cite{SME} (for more details, see also \cite{Mat}). Moreover, if we require that the theory at high temperature is invariant under chiral transformations, these terms are ruled out, since $\{\gamma_5,e^\mu + if^\mu\gamma_5 + \textstyle{1\over 2}g^{\lambda\nu\mu}\sigma_{\lambda\nu}\}\neq0$.

Therefore, the remaining coefficients are $c_{\mu\nu}$, $d_{\mu\nu}$, and $(k_F)_{\mu\nu\lambda\rho}$. Now, in order to get a Clifford algebra for $\Gamma^\mu = \gamma^\mu + c^{\mu\nu}\gamma_\nu + d^{\mu\nu}\gamma_5\gamma_\nu$, we must choose $d_{\mu\nu}=Q(\delta_{\mu\nu}+c_{\mu\nu})$, where $Q$ is a constant \cite{Aria}. With this assumptions, the theory (\ref{Lbfe}) becomes
\ba\label{Lbfe2}
{\cal L}_\mathrm{high}&=&-\frac{1}{4}F_{\mu\nu}F^{\mu\nu}-\frac14 (k_F)_{\mu\nu\lambda\rho}F^{\mu\nu}F^{\lambda\rho}+\bar{\psi}\left[i\partial_\mu(g^{\mu\nu}+c^{\mu\nu})\tilde\gamma_\nu-e\,A_\mu(g^{\mu\nu}+c^{\mu\nu})\tilde\gamma_\nu\right]\psi \nonumber\\
&& + {\cal L}_\mathrm{gf} + {\cal L}_\mathrm{gh},
\ea
where $\tilde\gamma^\mu=(1+Q\gamma_5)\gamma^\mu$.

\begin{figure}[ht]
\centering
\includegraphics{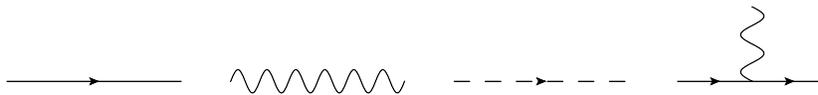}
\caption{Feynman Rules. Continuous, wavy, and dashed lines represent the fermion, photon, and ghost propagators, respectively, with momenta $\tilde p^\mu=((1+\kappa)p^0,p^i)$. The fermion-photon vertex is the usual one, $-ie\,\tilde\gamma^\mu$.}
\label{FR}
\end{figure}

We now assume rotational invariance, such that the coefficients in~(\ref{Lbfe2}) may be reduced to products of a given unit timelike vector $u_\mu$, which describes the preferred frame (see \cite{JacLib}, for more details). Proceeding in this way, we write $c_{\mu\nu}= \kappa\,u_\mu u_\nu$, and 
\be
(k_F)_{\mu\nu\lambda\rho} = \tilde\kappa\,(g_{\mu\lambda}u_\nu u_\rho + g_{\nu\rho}u_\mu u_\lambda - g_{\mu\rho}u_\nu u_\lambda - g_{\nu\lambda}u_\mu u_\rho),
\en
see Eq.~(\ref{kF}), where $u_\mu=(1,0,0,0)$ and now $\kappa$ and $\tilde\kappa$ are the coefficients that determine the scale of Lorentz violation. By choosing $\alpha=1$ (Feynman gauge) and $\tilde\kappa=(1+\frac{\kappa}{2})\kappa$, which is convenient to keep the Lagrangian (\ref{Lbfe2}) formally covariant, we get 
\be
{\cal L}=-\frac{1}{4}\tilde F_{\mu\nu}\tilde F^{\mu\nu} +i\bar{\psi}\tilde\partial_\mu\tilde\gamma^\mu\psi-e\,\bar\psi\tilde A_\mu\tilde\gamma^\mu\psi +\frac12(\tilde\partial_\mu \tilde A^\mu)^2 + (\tilde\partial_\mu\bar C)(\tilde\partial^\mu C),
\en
where $\tilde F_{\mu\nu}=\tilde\partial_\mu\tilde A_\nu - \tilde\partial_\nu\tilde A_\mu$, with $\tilde\partial_\mu=((1+\kappa)\partial_0,\partial_i)$ and $\tilde A^\mu=((1+\kappa)A^0,A^i)$.
The corresponding Feynman rules are shown in Fig.~\ref{FR}.

\section{The free energy}\label{fe}

Let us now compute the free energy per unit of volume (pressure) as a function of the temperature $T$ and of the coupling constant $e$, in the regime of high temperature. We shall calculate the expression for the pressure to order $e^2$, which has the form 
\be\label{P}
P = \frac{T\ln Z}{V} = P_0 + P_2,
\en
where $P_0$ is the zero order contribution in the coupling constant and $P_2$ is the second order one. The free energy density is minus the expression (\ref{P}).  

\subsection{One-loop contribution}

\begin{figure}[ht]
\centering
\hspace{-0.5cm}
\includegraphics{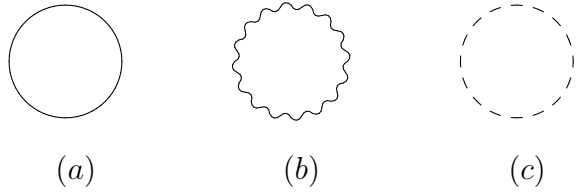}
\caption{One loop vacuum diagrams.}
\label{oneloop}
\end{figure}

The lowest-order contributions are given by the three one-loop vacuum diagrams, displayed in Fig.~\ref{oneloop}, and written in the imaginary time formalism as 
\be\label{P0}
P_0=\tr\sumint \{dp\}\ln\Slash{\tilde p}+4(-{\textstyle\frac12})\,\sumint dp\ln\tilde p^2+2({\textstyle\frac12})\,\sumint dp\ln\tilde p^2, 
\en
respectively, where we have introduced the shorthand notation
\be
\sumint \{dp\}=T\sum_{p_0=2\pi (n+\frac12)T}\int\frac{d^3p}{(2\pi)^3}
\en
for fermionic loop momenta and
\be
\sumint dp=T\sum_{p_0=2\pi nT}\int\frac{d^3p}{(2\pi)^3}
\en
for bosonic loop momenta, and $\Slash{\tilde p}=\tilde p_\mu\tilde\gamma^\mu$.

As usual, the bosonic contributions of Eq.~(\ref{P0}) have four degrees of freedom for the gauge field (Fig.~\ref{oneloop}(b)) and two degrees of freedom for the ghost field (Fig.~\ref{oneloop}(c)). 
The fermionic contribution, after the calculation of the trace,
\be
P_0^f=2\sumint \{dp\} \ln[(1-Q^2)(p_0^2(1+\kappa)^2+{\bf p}^2)],
\en
has the same form as the bosonic contributions, 
\be
P_0^b=-\sumint dp \ln(p_0^2(1+\kappa)^2+{\bf p}^2).
\en
In order to evaluate these expressions we proceed similarly to \cite{DJ},  such that
\be
P_0=\int\frac{d^3p}{(2\pi)^3}\left[\frac{|{\bf p}|}{1+\kappa}+\frac{4}{\beta}\ln\left(1+e^{-\frac{|{\bf p}|\beta}{1+\kappa}}\right)-\frac{2}{\beta}\ln\left(1-e^{-\frac{|{\bf p}|\beta}{1+\kappa}}\right)\right],
\en
which is independent of the parameter $Q$. The zero temperature (divergent) part is absorbed in a renormalization of the vacuum energy \cite{Kapusta}. Therefore, after integrating in the angular variables, we get
\be
P_0=\frac{2}{\pi^2\beta}\int_0^{\infty}dr\,r^2\ln\left(1+e^{-\frac{r\beta}{1+\kappa}}\right)-\frac{1}{\pi^2\beta}\int_0^{\infty}dr\,r^2\ln\left(1-e^{-\frac{r\beta}{1+\kappa}}\right),
\en
and finally, we obtain 
\be
P_0=\frac{11\pi^2}{180}T^4(1+\kappa)^3.
\en

Therefore we conclude that the only impact of the Lorentz symmetry breaking in the case of the CPT-even Lorentz-breaking parameter consists in the modification of a constant factor.

\subsection{Two-loop contribution} 

\begin{figure}[ht]
\includegraphics{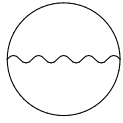}
\caption{Two-loop diagram.}
\label{twoloop}
\end{figure}

In this subsection we study the two-loop contributions to the free energy. The Lorentz violating contribution to free energy up to the two-loop order (which corresponds to the second order in $e$, is given by Fig.~\ref{twoloop}), 
which can be written as
\be\label{P2}
P_2 =\frac12 e^2\sumint \{dp\}\,\sumint \{dq\}\,\tr\,\left[\tilde\gamma^\mu\frac{1}{\Slash{\tilde p}}\tilde\gamma_\mu \frac{1}{\Slash{\tilde q}}\frac{1}{(\tilde p+\tilde q)^2}\right] . 
\en
Note that, due to the fact that
\be
\tilde\gamma^\mu\frac{1}{\tilde p_\alpha\tilde\gamma^\alpha}=\gamma^\mu\frac{1}{\tilde p_\alpha\gamma^\alpha},
\en
the above second order contribution is also independent of the parameter $Q$, and so on for the other contributions. Thus, we can rewrite (\ref{P2}) as
\be
P_2 =\frac12 e^2\sumint \{dp\}\,\sumint \{dq\}\,\tr\,\left[\gamma^\mu \frac{\tilde p_\alpha\gamma^\alpha}{\tilde p^2}\gamma_\mu \frac{\tilde q_\beta\gamma^\beta}{\tilde q^2}\frac{1}{(\tilde p+\tilde q)^2}\right],
\en
so that, after calculating the trace, we get
\be
P_2 = 2e^2\sumint \{dp\}\,\sumint\{dq\}\left[-\frac{1}{\tilde p^2\tilde q^2}+\frac{1}{\tilde q^2(\tilde p+\tilde q)^2}+\frac{1}{\tilde p^2(\tilde p+\tilde q)^2}\right].
\en

Using the results for the sum-integrals,
\ba
\sumint \{dp\}\,\sumint\{dq\}\frac{1}{\tilde p^2\tilde q^2}&=&\frac{1}{576}T^4(1+\kappa)^2,\\
\sumint \{dp\}\,\sumint\{dq\}\frac{1}{\tilde q^2(\tilde p+\tilde q)^2}&=&-\frac{1}{288}T^4(1+\kappa)^2,
\ea
we arrive at the following Lorentz violating contribution to the free energy,
\be
\label{res}
P_2 = -\frac{5}{288}e^2T^4(1+\kappa)^2.
\en
Thus, the modification in the free energy due to the Lorentz-breaking parameter for the two-loop order consists only in multiplying by a constant, just as in the one-loop order. 

Therefore, the expression for the pressure to order $e^2$ in the presence of Lorentz symmetry breaking looks like
\ba\label{P}
P=\frac{\pi^2T^4}{9}\left[ \frac{11}{20}(1+\kappa)^3-\frac{5e^2}{32\pi^2}(1+\kappa)^2\right].
\ea

\section{Numerical estimations}\label{ne}

In order to estimate a bound for the Lorentz violating parameter $\kappa$, we use the theoretical predictions of the primordial helium abundance $Y$ developed in the references \cite{Dicus, Heck, Lopez}. A way to determine $Y$ consists in the analysis of the change in thermodynamics quantities such as the energy density $\rho$, the pressure $P$ and the neutrino temperature $T_\nu$. Using the result (\ref{P}), let us now study these changes in the presence of the Lorentz-breaking.

The contribution to the energy density can be found from the standard thermodynamic relation $\rho=-P+T(\partial P/\partial T)$, so that
\be
\rho = \frac{\pi^2T^4}{15} (N+\delta N),
\en
where $N=\frac{11}{4}$ and $\delta N\approx-0.007+8.236\kappa$. Thus, we obtain
\be
\frac{\Delta\rho}{\rho} \approx -2.5\times10^{-3}+2.994\kappa.
\en
The fraction $\Delta Y$ is affected by QED in several ways, as can be seen in \cite{Heck}. The total effect is approximately 
\be
\Delta Y \approx 2.9\times10^{-4} + 0.15\frac{\Delta T_\nu}{T_\nu} + 0.07\frac{\Delta\rho}{\rho},
\en
where $T_\nu$ depends on energy density $\rho$ (for more details, see \cite{Wein}).

The usual theoretical result, without parameter $\kappa$, is $\Delta Y\approx10^{-4}$, whereas the experimental one is $\Delta Y\approx10^{-3}$ \cite{Ioc}. Therefore, an upper bound for $\kappa$ necessary for the coincidence of the theoretical and experimental results must be $\kappa\sim10^{-2}$. This result agrees with the value found in \cite{Lane,Russell}.

\section{Summary}\label{su}

We have calculated the contributions to the free energy in the rotationally invariant Lorentz-violating QED in one- and two-loop approximations at high temperature. The corresponding correction to the pressure was then determined.  

We also observe  that Lorentz violation can be used to explain the difference between theoretical and experimental predictions of the primordial helium abundance. By matching these predictions, we have estimated the Lorentz-breaking parameter $\kappa\sim10^{-2}$, which agrees with that obtained in \cite{Lane}.

{\bf Acknowledgements.} This work was partially supported by Conselho
Nacional de Desenvolvimento Cient\'{\i}fico e Tecnol\'{o}gico (CNPq)
and Funda\c{c}\~{a}o de Amparo \`{a} Pesquisa do Estado de S\~{a}o
Paulo (FAPESP), Coordena\c{c}\~{a}o de Aperfei\c{c}oamento do Pessoal
do Nivel Superior (CAPES: AUX-PE-PROCAD 579/2008) and CNPq/PRONEX/FAPESQ.

\end{document}